\documentstyle[preprint,aps,tighten,graphicx]{revtex}

\newcommand{\expl}{\langle \!\langle}
\newcommand{\expr}{\rangle \!\rangle}
\def\Journal#1#2#3#4{{#1} {\bf #2}, #3 (#4)}

\def\NPA{{\em Nucl.~Phys.} A}
\def\PLB{{\em Phys.~Lett.}  B}
\def\PRL{\em Phys.~Rev.~Lett.~}

\def\PRC{{\em Phys.~Rev.} C}
\def\ZPC{{\em Z.~Phys.} A}
\def\ZPC{{\em Z.~Phys.} C}

\def\PRep{\em Phys.~Rep.~}

\def\JP{{\em J.~Phys.} G}

\begin{document}

\title{Importance of multi-mesonic fusion processes
on (strange) antibaryon production}

\author{C.~GREINER}
\address{Institut f\"ur Theoretische Physik, Universit\"at Giessen,
   Heinrich-Buff-Ring 16, D-35392 Giessen, Germany}


\maketitle

\begin{abstract}
Sufficiently fast chemical equilibration of
(strange) antibaryons in an environment of nucleons, pions and kaons
during the course of a relativistic heavy ion collision
can be understood by a `clustering' of
mesons to buildt up baryon-antibaryon pairs.
This multi-mesonic (fusion-type) process
has to exist in medium due to the principle of detailed balance.
Novel numerical calculations for a dynamical setup are
presented. They show that
- at maximum SPS energies - yields
of each antihyperon specie are obtained which are consistent
with chemical saturated populations of $T \approx  150-160 $ MeV,
in line with popular chemical freeze-out parameters extracted
from thermal model analyses.

\end{abstract}


\section{Brief Overview on Antihyperon Production}

Strangeness enhancement
has been predicted  a long time ago as a
potential probe to find clear evidence for the temporary existence of a
quark gluon plasma (QGP) in relativistic heavy ion collisions.
A strong experimental effort has been made since
and is still made
for measuring strange particle abundancies in
experiments at Brookhaven and at CERN
(for a short recent review see \cite{Stock}).
In particular because of high production thresholds in
binary hadronic reaction channels
antihyperons had been advocated as the
appropriate QGP candidates \cite{KMR86}.
Indeed, a satisfactory picture
of nearly chemically saturated populations of
antihyperons has been experimentally demonstrated over the
last years
with the Pb+Pb experiments NA49 and WA97 at CERN-SPS.
For this statement, of course, a quantitative,
theoretical analysis by employing a thermal (or `statistical') model
has to be invoked by fitting the thermodynamical parameters
to the set of individual (strange) hadronic abundancies \cite{BMS96}.

On the other hand, already since the first measurements
with the lighter ions have been undertaken in the early
nineties, the theoretical description of the antibaryon
production within hadronic transport schemes
in comparison to these data
faced some severe difficulties.
Some phenomenological motivated attempts
to explain a more abundant production of antihyperons within
a hadronic transport description \cite{So95} had been
proposed like the appearance of color ropes,
the fusion of strings,
the percolation of strings, or the formation of high-dense
hadronic clusters.
The underlying mechanisms, however, have to be considered
exotic and to some extent ad hoc, their purpose lies mainly
to create (much) more antibaryon in the very early
intial stage of the reaction (compared to simple
rescaled p+p collisions). To some extent indeed the
philosophy behind these mechanisms was to model a precursor of
initial QGP formation within a hadronic transport scheme.
On the other hand,
in most of the transport calculations
a dramatic role of subsequent antibaryon annihilation
is observed, which, in return, has to be more than counterbalanced
by these more exotic, initially occuring mechanisms.
Large annihilation rates
result if the free-space cross section is employed.
The situation seems even more paradox with respect to the fact that
the chemical description within thermal models works indeed
amazingly well
for the antihyperons and being, of course, completely nondependent on
the (large) magnitude of the annihilation cross section.
For all of this reasons the
theoretical and dynamical understanding of the
production of (strange) antibaryons has remained a delicate and challenging
task \cite{So98}.

\begin{figure}
\centering
   \includegraphics[height=6cm]{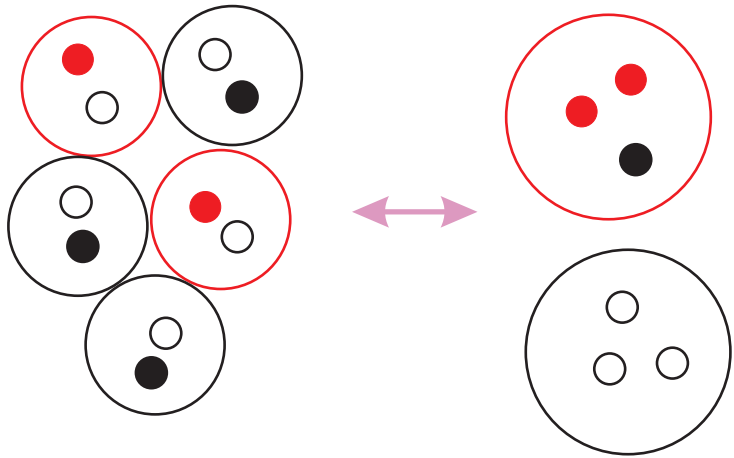}
\centering
\caption{Schematic picture for the multi-mesonic fusion-like
reaction $3 \pi  + 2  K \leftrightarrow \bar{\Xi } + N$.}
\label{fig:Fusion}
\end{figure}

As we will demonstrate,
a correct incorporation of the baryonic annihilation
channels had actually not been done consistently.
We have conjectured recently that
a sufficiently fast redistributions of strange and light quarks
into (strange) baryon-antibaryon pairs should be achieved
by multi-mesonic fusion-type reactions of the type
\begin{equation}
\label{mesfuse}
n_1\pi + n_2 K \, \leftrightarrow \, \bar{Y}+p
\end{equation}
occuring in a moderately dense hadronic system \cite{GL00}
(illustrated in Fig.~1).
The beauty of this argument lies in the fact that
(at least) these special kind of multi-hadronic
reactions have to be present because of the fundamental principle of detailed
balance.
As the annihilation of antihyperons on baryons is of dramatic relevance,
the multi-mesonic (fusion-like) `back-reactions'
involving $n_1$ pions and $n_2$ kaons, where $n_2$ counts
the number of anti-strange quarks within the antihyperon $\bar{Y}$),
must, in principle, be taken care of in a dynamical simulation.
These reactions are then the most
dominant source of production.
This crucial fact had been overseen in all of the aforementioned
treatments.
The underlying reasoning was first raised by Rapp and Shuryak
who described the maintenance of nearly perfect chemical
equilibrium of antiprotons together
with pions and nucleons during the very late stage of the expanding
fireball before the particles loose contact \cite{RS00}.

The crucial input, though plausible, is now to assume that the
annihilation cross sections for any strange or nonstrange antibaryon
on any baryon
are approximately the same order as for $N\bar{p}$
at the same relative momenta, i.e.
$\sigma _{B \bar{Y}\rightarrow n  \pi + n_Y K} \approx
\sigma _{N \bar{p}\rightarrow n  \pi } $,
being in the range of 50-100 mb for characteristic and
moderately low momenta occuring in an expanding hadronic fireball.
The equilibration timescale
$(\Gamma _{\bar{Y}})^{(-1)} \sim
1/ (\sigma _{B \bar{Y}} v _{B\bar{Y}} \rho_B )$
is to a good approximation proportional to the
inverse of the density of baryons and their resonances.
Adopting an initial density of 1--2 times
normal nuclear matter density $\rho_0 $ for the initial and thermalized
hadronic fireball, the antihyperons do equilibrate
on a timescale of 1--3 fm/c well
within the expansion timescale of the late hadronic fireball.
Hence, fast chemical equilibration
of the antihyperon abundancies is guaranteed by detailed
balance with respect to the strong annihilation, the final
yields then being independent of the actual size
of the (large) annihilation cross section, solving the
aforementioned paradox.

To be quantitative,
(some) novel results by solving rate calculations
for a dynamical setup are presented
in the following. Before turning to their discussion,
we will review briefly on some general ideas
of the baryon-antibaryon annihilation process,
in order to strengthen the one main assumption
concerning the general size of its cross section, and
on how to come to general master equations.

\section{Antibaryon Annihilation and its Effective Description
by a Master Equation}

Antinucleon-nucleon annihilation is the strongest of all
strong interaction processes.
The strong annihilation of a nucleon and an antinucleon
can be thought quantum mechanically
as a complete absorbtive
scattering process given approximately by the black disk formula
$\sigma_{abs} = \pi (R+\lambda )^2$ or by a more sophisticated boundary
condition description \cite{JV86}.
For $p+\bar{p}$-annihilation one finds for the
`black disk'  radius $R=1.07$ fm, for which then
this decription reproduces very accurately the total inelastic cross section
as a function of the beam momentum \cite{JV86} and also
its steep increase and diverging behaviour at low momenta.
Hence, the quantum mechanical interpretation of the annihilation
(being exotherm) is a picture of complete absorption.
One is now tempted to
generalize this simple and intuitive picture for {\em } all
baryon-antibaryon annihilation processes. The only `free' parameter
which then can change is the radius $R$. Again, as it
basically reflects the radius of the proton in the case of
$p\bar{p}$-annihilation, one would expect that
this canonical value of $1$ fm is of general validity
for all the annihilation processes. Thus it is plausible
to assume that the cross section for annihilation between
any baryon and antibaryon should be rather the same for
the same relative momenta.
Indeed, there does exist old data on the total $\bar{\Lambda }+p$
cross section \cite{E76} with reasonably large value, although
the $\bar{\Lambda }$ momenta was exceeding $4$ GeV/c in all data taken.
(These data are not conclusive to really extrapolate
dwn to lower momenta, though they give a right indication:
The authors of \cite{E76} reported an extrapolated formula of
$\sigma_{abs} = 47 (\pm 10) p_{\Lambda }^{-1/2} $ mb$\, $GeV$^{-1/2}$,
which is not fully analogue to the black disk formula,
but has already a rather correct shape with a sizeable
magnitude.)

A more microscopic and quite popular picture, which can decribe quantitatively
the complicated final states of
individual outgoing mesonic channels, is
the concept of two meson doorway states \cite{JV88} given by
\begin{equation}
\label{twomes}
\bar{B} + B \, \rightarrow \, M_1 + M_2 \rightarrow
\, n_1\pi + n_2 K \, \, .
\end{equation}
The main assumption is that the annihilation occurs exclusively
via two mesons with nearest threshold dominance.
$M_1$ and $M_2$ can be rather highlying resonances and/or hybrid
states which come close with their individual threshold.
The final decay of these two mesons
into the various channels of multiple pions and kaons is then
treated statistically and microcanonically.
It turns out that such a description describes rather nicely
reproduces the available data and can also be generalized
straightforwardly for antihyperon annihilation \cite{JV88}.
There do exist a lot of other phenomenological
and microscopic descriptions like quark models or flux tube models,
which try to describe quantitatively the complicated process
of the annihilation in more dynamical terms (for a review see \cite{D92}).

How can one incorporate such complicated processes into a description
of transport dynamics? It is clear that as the very microscopy of the
annihilation processes is not fully understood, one has to
abandon any detailed local modelling, but has to turn to an effective
coarse grained description which is guided by physical principles.
Following the concepts of relativistic kinetic theory, the
microscopic starting point is
a Boltzmann-type equation of the form
{\footnotesize
\begin{eqnarray}
\partial_t f_{\bar{Y}} + \frac{{\bf p}}{E_{\bar{Y}}} {\bf \nabla } f_{\bar{Y}}
& = &
\sum_{ \{ n_1 \} ; B} \frac{1}{2E_{\bar{Y}}}
\int \frac{d^3p_B}{(2\pi )^3 2E_B}
\prod _{ \{ n_1, n_2 \} } \int \frac{d^3p_i}{(2\pi )^3 2E_i}
\, (2\pi )^4
\delta^4 ( p_{\bar{Y}} + p_B - \sum_{\{ n_1, n_2 \} }p_i )
\nonumber \\
&&
| \expl n_1,n_2 | T | \bar{Y} B \expr |^2
\left\{ (-) f_{\bar{Y}}f_B \prod_{\{n_1, n_2 \} } (1+f_i) \, + \,
\prod _{ \{ n_1, n_2 \} } f_i
(1- f_{\bar{Y}}) (1-f_B) \right\} \, \, ,
\nonumber
\end{eqnarray}
}
where
{\footnotesize
\begin{eqnarray}
\label{cross}
\sigma _{\bar{Y}B}^{\{n_1 \} } & \equiv &
\frac{1}{`Flux'}
\sum_{\{ n_1 \}}
\prod _{ \{ n_1, n_2 \} } \int \frac{d^3p_i}{(2\pi )^3 2E_i}
\, (2\pi )^4
\delta^4 \left( p_{\bar{Y}} + p_B - \sum_{ \{ n_1, n_2 \} }p_i \right)
| \expl n_1,n_2 | T | \bar{Y} B \expr |^2
\nonumber
\end{eqnarray}
}
corresponds to the total annihilation cross section
into the various possible multiple pion states.
This form of the collision exactly incorporates the principle of detailed
balance. The static fixed point of this equation
(together with the usual binary kinetic processes)
are thermal and chemical saturated distributions.
Without the back reaction channels the equations would only have
vanishing distributons as stable fixed points, which is, of course,
unacceptable. Hence, the back reactions have to be considered as
very basic and important ingredient of the transport description when
trying to implement the baryon-antibaryon annihilation processes.

Furthermore it is obvious that the back reactions will
guarantee that the (strange) antibaryons
become chemical saturated with the pions, kaons and nucleons
on a very short timescale. To see this more explicit, and also
for the further numerical treatment, one can bring the above
Boltzmann equation into a more intuitive form of a master or rate
equation.
Assuming $v_{rel}\, \sigma_{\bar{Y}B} (\sqrt{s} ) $ to be roughly constant,
which is actually
a good approximation for the $p\bar{p}$-annihilation,
or, invoking a standard description of further effective
coarse graining by using
thermally averaged cross sections
and distributions,
and furthermore
taking the distributions in the Boltzmann approximantion,
the following master equation
for the respectively considered antihyperon density
is obtained
\begin{equation}
\label{mastera}
\frac{d}{dt} \rho _{\bar{Y}} \, =\,   -
\expl \sigma _{\bar{Y}B} v _{\bar{Y}B} \expr
\left\{
\rho _{\bar{Y}} \rho_B \,  \vphantom{\sum_{n}}
-  \, \sum_{\{ n_1 \} }
\hat{M}_{(n_1,n_2)}(T,\mu_B,\mu_s) (\rho _\pi)^{{n_1}} (\rho _K )^{n_2}
\right\},
\end{equation}
where the
`back-reactions' of several effectively clustering
pions and kaons are incorporated in the `mass-law' factor
$$
\hat{M}_{(n_1,n_2)}(T,\mu_B,\mu_s) \, = \,
\frac{ \rho _{\bar{Y}}^{eq.} \rho ^{eq.}_B }
{(\rho ^{eq.}_\pi)^{{n_1}} (\rho ^{eq.}_K )^{n_2}} \, p_{n_1} \, \, \, .
$$
Here $p_{n_1}$ (which will generally depend on the
thermodynamical parameters) states the relative probability
of the reaction (\ref{mesfuse}) to decay into a specific number $n_1$ of pions
and $\rho_B$ denotes the total number density of baryonic particles.
As is well-known, the mass-law factor
$\hat{M}$ depends only on the temperature and the baryon and strange quark chemical
potentials.
$\Gamma  _{\bar{Y}} \equiv
\expl \sigma _{\bar{Y}N} v _{\bar{Y}N} \expr \rho_B $
gives the effective annihilation rate of the respective antihyperon
specie on a baryon.

\begin{figure}
\centering
   \includegraphics[height=8cm]{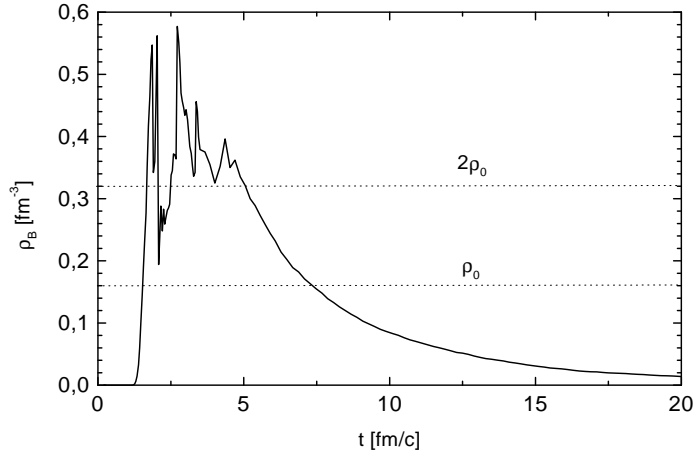}\\
\centering
    \caption{
     Time evolution of the (average) net baryon density
for midrapidity $|\Delta Y| \leq 1$ and central Pb+Pb-collision
at 160 AGeV obtained within a dynamical transport simulation.
Here the amount of baryon number residing still in string-like
excitations is explicitely discarded.
String-like excitations have disappeared after
about 3-4 fm/c, so that
from this time on a pure hadronic fireball develops and expands.
Its initial net baryon-density starts slightly above $\rho_B=2\rho_0$.
}
\label{Cassing}
\end{figure}

For a further manipulation one has to make assumptions for the
various abundancies occuring in the master equation.
Nonequilibrium inelastic hadronic reactions
can explain to a good extent the overall strangeness production
seen experimentally:
The major amount of the produced kaons at SPS-energies
can be understood in terms of still early and energetic
non-equilibrium interactions \cite{Ge98}.
In Fig.~2 the time evolution of netbaryon density at midrapidiy
obtained within a dynamical transport simulation \cite{WC}
is depicted. All strangeness is being produced
still when string-like excitations are governing the dynamics.
In the later hadronic stage the number of strange quarks
stays more or less constant and can only be redistributed
among the various hadrons \cite{CG01}. Also the pions and nucleons
do stay more or less at thermal equilibrium.
Refering to the master equation (\ref{mastera}),
one can then take the pions, baryons and kaons to stay approximately
in thermal equilibrium throughout the later hadronic evolution of the collision,
the later being modelled to be an isentropic expansion
with fixed total entropy content being
specified via the entropy per baryon ratio $S/A$
(compare with Fig.~3).
(\ref{mastera}) has the intuitive form
\begin{equation}
\label{masterd}
\frac{d}{dt} \rho _{\bar{Y}} \,  = \,    - \,
\Gamma  _{\bar{Y}}
\left\{
\rho _{\bar{Y}} \, -  \,
\rho ^{eq }_{\bar{Y}}
\right\} \, \, \, .
\end{equation}
The production rate per unit volume
$dN_{\bar{Y}}/dtdV$ is given by
the non-vanishing, though small value $\Gamma_{\bar{Y}}\rho_{\bar{Y}}^{eq}$.
Still this rate is enough to populate the antihyperons to their
(very small) equilibrium value.

\begin{figure}
\centering
   \includegraphics[height=8cm]{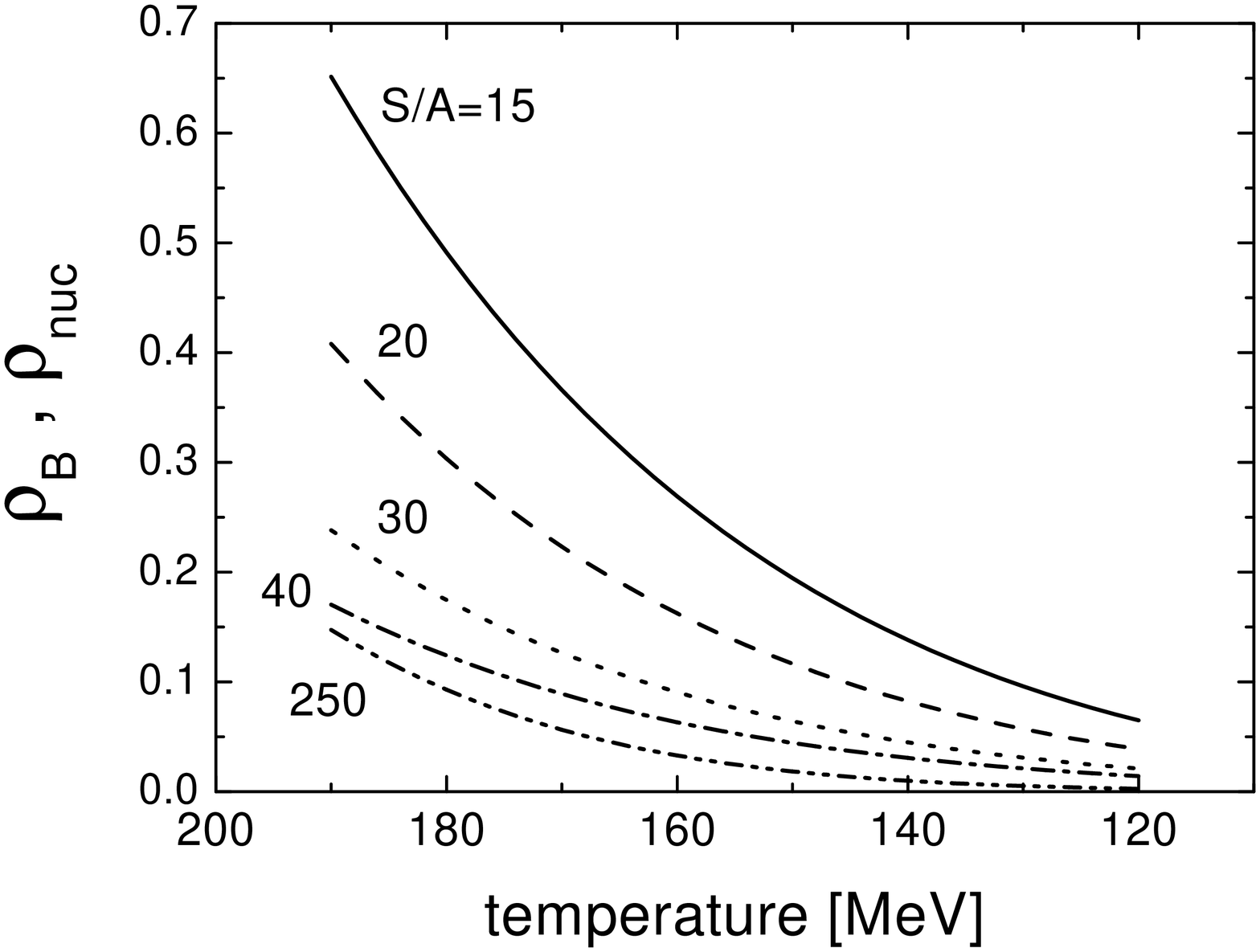}\\
\centering
    \caption{
The netbaryon density $\rho_B$ of a chemically fully equilibrated
thermal hadronic resonance gas
as function of decreasing temperature
for the situation of an isentropic expansion, i.e.
for various constant entropy/baryon ratios.
For the value $S/A=250$ (a situation expected at RHIC) the
total baryon number density $\rho_{nuc} $ is plotted instead.
}
\label{Isentrope}
\end{figure}
For the calculations solving this master equation (\ref{masterd})
one has to employ an `effective' volume $V(t)$
in order to extrapolate from Fig.~2 to Fig.~3, i.e.
simulating the global characteristic of the expansion and the
dilution of the baryon density and thus the annihilation rate
$\Gamma _{\bar{Y}}(t)$.
The `effective' (global or at midrapidity) volume $V(t)$
is parametrized as function of time by longitudinal
Bjorken expansion and including
a transversal expansion
either with a linear profile
\begin{equation}
\label{volumea}
V_{eff,lin}(t\geq t_0) \,  =  \, \pi \, (ct) \,
\left(R_0 + v_{lin}(t-t_0)  \right)^2
\end{equation}
(taking $v_{lin}$ as an appropriate parameter to simulate
slower or faster expansion)
or with
an accelerating
radial flow
\begin{equation}
\label{volume}
V_{eff,acc}(t\geq t_0) \, = \, \pi \, (ct) \,
\left(R_0 + v_0(t-t_0) + 0.5 a_0 (t-t_0)^2 \right)^2
\end{equation}
with $R_0 = 6.5\, fm$, $v_0= 0.15\, c$ and $a_0 = 0.05 \, c^2/fm$
for the later modeling.

\section{Results and Implications}

\begin{figure}
\centering
   \includegraphics[height=85mm]{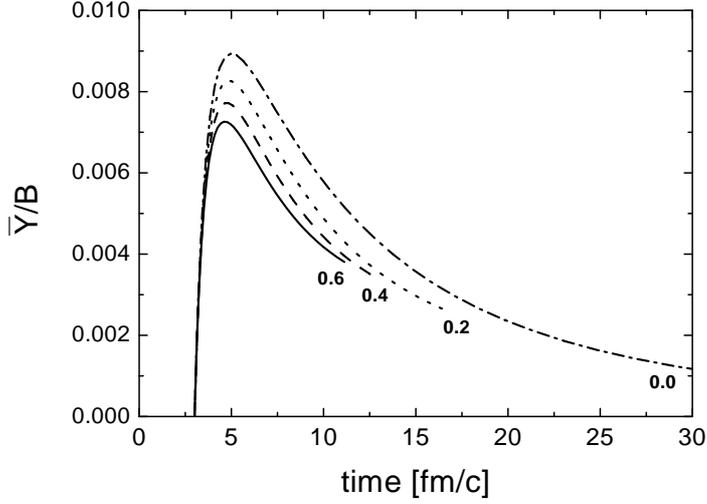}\\
\centering
    \caption{
        The anti-$\Lambda $ to baryon number ratio
$N_{\bar{ \Lambda }}/N_B \, (t)$
as a function of time for various velocity
parameters $v_{lin}$ for the transverse expansion.
The entropy per baryon is taken as $S/A=30$, $t_0=3 $ fm/c and
$T_0=190 $ MeV.
}
\end{figure}

At starting time $t_0 $ an initial temperature $T_0$ is chosen.
($T_0$ is set to $190$ MeV for the SPS
and $150 MeV$ for the AGS situation, while
the initial energy densities are
then about 1 GeV/fm$^3$.)
From (\ref{volumea}) or (\ref{volume}) together
with the constraint of conserved entropy
the temperature and the chemical potentials do
follow as function of time and thus also $\rho_B(t)$ as well as
$\rho_{\bar{Y}}^{eq}(t)$ within the hadronic resonance gas.
As a {\em minimal } assumption the initial abundancy of antihyperons
is set to zero. Equation (\ref{masterd}), taking into account
the volume dilution, is solved for each specie.
For the thermally averaged cross section we take a simple constant value
of $\langle \sigma_{ann} v \rangle := \sigma_0  \equiv 40$ mb.

\begin{figure}
\centering
   \includegraphics[height=85mm]{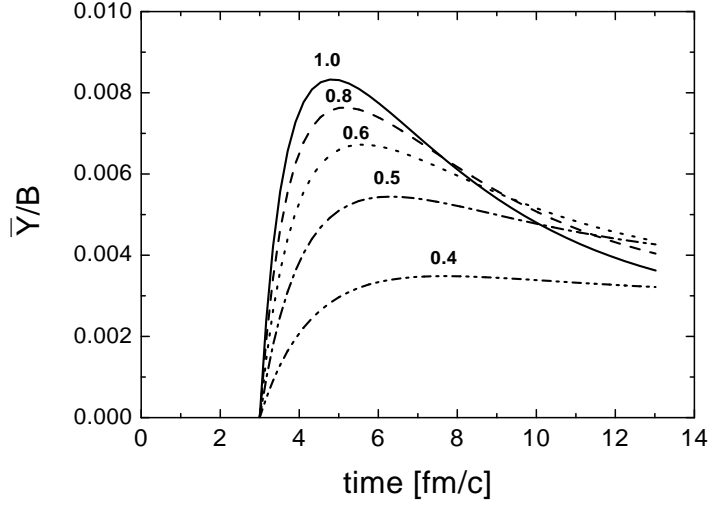}\\
\centering
    \caption{
        The anti-$\Lambda $ to baryon number ratio
$N_{\bar{ \Lambda }}/N_B \, (t)$
as a function of time for various
implemented annihiation cross section $\sigma_{eff} \equiv \lambda \,
\sigma_0 $.
The entropy per baryon is taken as $S/A=30$, $t_0=3 $ fm/c and
$T_0=190 $ MeV.
}
\end{figure}
In Fig.~4 the number of $\bar{\Lambda }$s
(normalized to the conserved net baryon number)
as a function of time is depicted.
The entropy per baryon is chosen as $S/A=30$ being
characteristic
to global (`$4\pi $') SPS results \cite{Cleymans}.
Here we have chosen the ansatz (\ref{volumea})for the expansion of the volume
being linear in time in the transverse direction.
The parameter $v_{lin}$ is varied to simulate slow or fast
expansion of the late hadronic fireball.
The general characteristics is that first the antihyperons
are dramatically being populated, and then in the very late
expansion some more are still being annihilated, depending on
how fast the expansion goes. A rapid expansion gives a higher yield,
which can increase the final yield by a factor of 2 to 3.
However, the typical expansion behaviour obtained
from simulations or extracted from the analysis of transverse
momentum slopes of individual hadrons (pions and protons) is that
at the late stages the transverse expansion velocity
shoul be about $0.5 $ c. In the following we stay to the second ansatz
(\ref{volume}) which extrapolates from an initially slower
to a later faster expansion.

\begin{figure}
\centering
   \includegraphics[height=11cm]{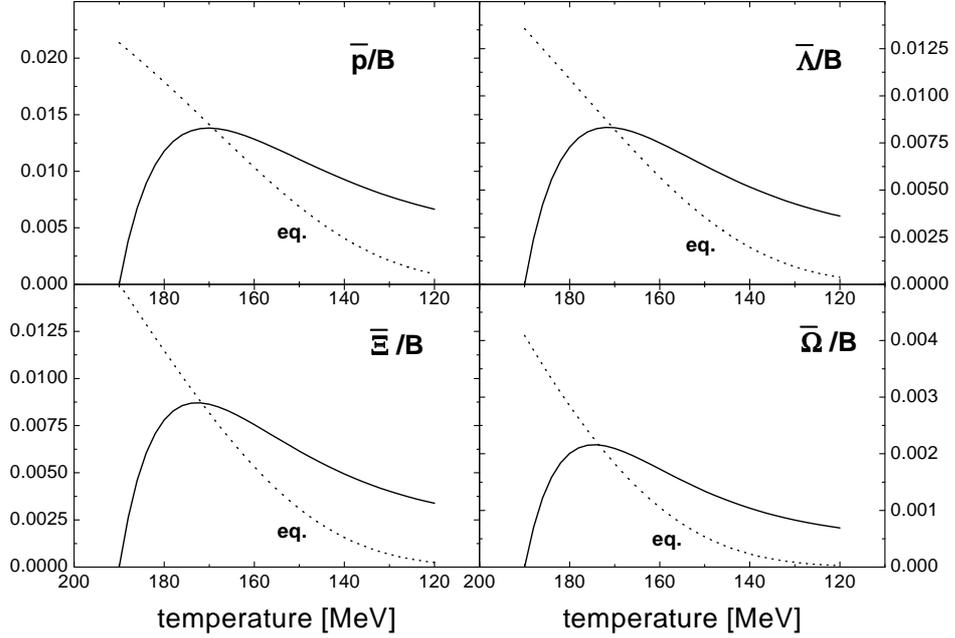}\\
\centering
  \caption{
  The antihyperon to baryon number ratio
$N_{\bar{Y}}/N_B \,  (T)$ and $N_{\bar{Y}}^{eq.}/N_B (T) \,  $
(dotted line) as a function of the decreasing temperature.
Parameters are the same as in Fig. 5.
}
\end{figure}

In Fig.~5 the number of $\bar{\Lambda }$s
as a function of time is depicted, where now
the cross section employed
is varied by a constant factor, i.e.
$\sigma_{eff} \equiv \lambda \, \sigma_0$.
The results are rather robust against a variation by a factor of 2
in the cross section. Typically (for $\lambda =1$)
about more than 5 times
in number of antihyperons are created during the evolution compared
to the final number freezing out, thus reflecting the
fast back (`annihilation') and forth (`creation')
processes at work dictated by detailed balance.

\begin{figure}
\centering
   \includegraphics[height=90mm]{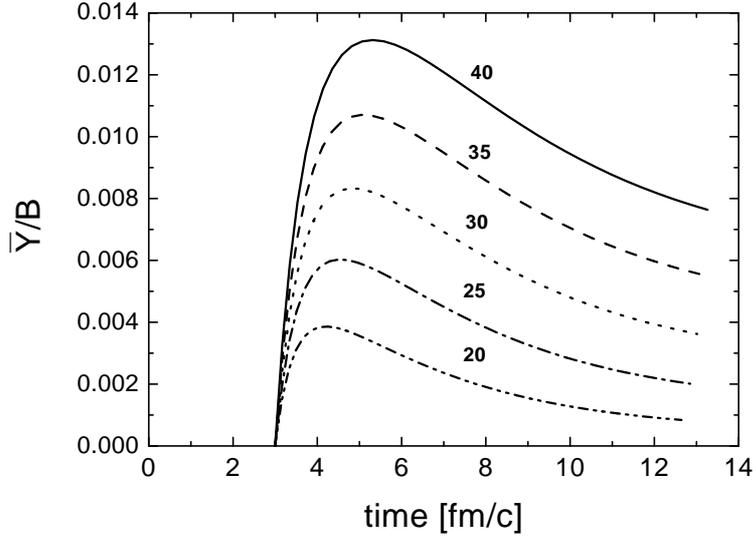}
\centering
       \caption{
      $N_{\bar{\Lambda }}/N_B \,(t) $ as a function of time
     for various entropy content described
  via the entropy per baryon ratio ($S/A=20-40$).
  Other parameters are as in Fig.~5.}
\end{figure}

In Fig.~6 the number of antihyperons of each specie
are now shown as a function of the decreasing temperature $T(t)$
of the hadronic system.
For a direct comparison the instantaneous equilibrium abundancy
$N_{\bar{Y}}^{eq.}(T(t),\mu_B(t),\mu_s(t))/N_B$ is also given.
As noted above, after a fast initial population,
the individual yields of the antihyperons do overshoot
their respective equilibrium number
and then do finally saturate at some slightly smaller value.
Moreover, one notices that
the yields effectively do saturate at a number
which can be compared to an equivalent equilibrium number
at a temperature parameter around $T_{eff} \approx 150-160$ MeV,
being strikingly close
to the ones obtained within the various thermal analyses \cite{BMS96}.

In Fig.~7 the number of anti-$\Lambda $s
as a function of time is given for various entropy per baryon ratios.
One notices that the final value in the yield significantly
depends on the entropy content, or, in other
words, on the baryochemical potential. We note that the results
at midrapidity from WA97 can best be reproduced by employing
an entropy to baryon ratio $S/A=40$. Indeed, at
midrapidity one qualitatively
expects a higher entropy content due to the
larger pion to baryon ratio as compared to
full `$4\pi$' data over all rapidities. At this point it will also
be very interesting to compare our semi-quantitative
calculations with the new results from NA49
on the $\bar{\Lambda }$-yield at lower SPS energies of 80
AGeV and 40 AGeV with lower entropy contents, respectively.
As the presented results are very sensitive
on the entropy content,
one first needs a clean analysis to obtain a rather
accurate $S/A$ number from the measured pion and proton abundancies.

\begin{figure}
\centering
   \includegraphics[height=90mm]{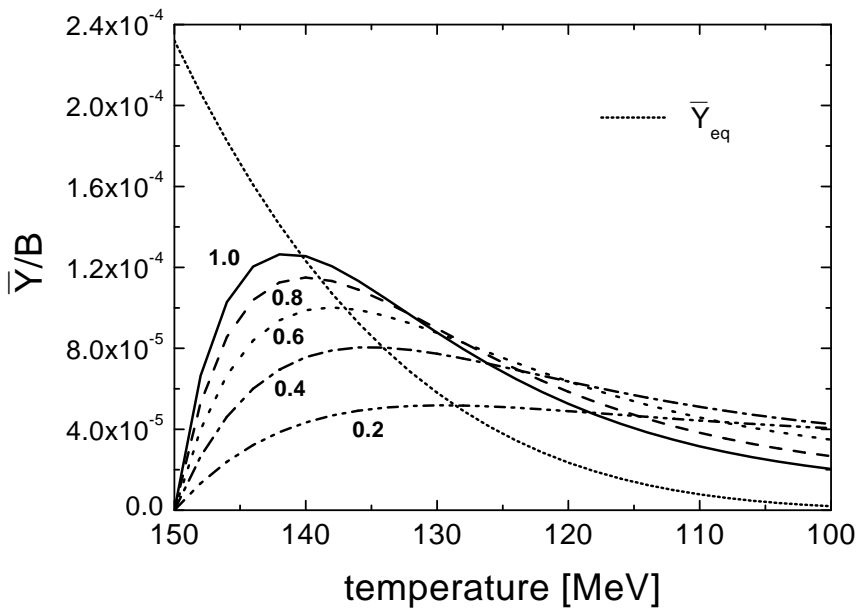}
\centering
        \caption{
      $N_{\bar{\Lambda }}/N_B \,(T) $
      and $N_{\bar{\Lambda }}^{eq.}/N_B (T) \,  $
      as a function of
      decreasing temperature for a characteristic AGS situation
     with an entropy content of $S/A=12$ for various
    implemented annihiation cross section
    $\sigma_{eff} \equiv \lambda \, \sigma_0 $.
   $t_0=5 $ fm/c and
   $T_0=150 $ MeV. }
\end{figure}

There is also a clear hint at AGS energies
of enhanced anti-$\Lambda $ production \cite{E802}.
For most central collisions more anti-$\Lambda $s are found
compared to anti-protons, which is quite puzzling as within
a thermal model analysis this ratio is found not to be larger
than 1. This enhanced ratio of anti-$\Lambda $s compared to
anti-protons at AGS energies one can
understand in a way that one assumes that their annihilation cross section
on baryons is just slightly smaller than for the antiprotons.
In Fig.~8 a similar study like that of Fig.~5 is shown for a characteristic
situation at AGS. For smaller, yet not too small effective cross sections
the final yield can here be enlarged by a factor of 2 compared to the
case with a `full' crossection, as the final reabsorption is not as effective.
But, of course, this idea is speculation at present.
Also, we remark that
the $\bar{\Lambda }$s effectively do saturate
at an equivalent equilibrium number
at a temperature parameter around $T_{eff} \approx 120-130$ MeV.
Unfortunately, there are no data for $\bar{\Xi }$ at AGS.
Again the new NA49 data at lower energies are worthwhile to pursue.
A detailed measurement of all antihyperons
represents also an excellent opportunity for future heavy ion facilities
at an energy upgraded GSI.

To summarize,
multi-mesonic production of antihyperons is a consequence
of detailed balance and, as the annihilation rate is
large, it is by far the most dominant source
in a hadronic gas.
This is a remarkable observation,
as it clearly demonstrates
the importance of hadronic multi-particle channels,
occuring frequently enough in a (moderately) dense {\em hadronic}
environment in order to populate and chemically saturate
the rare antibaryons.
In order to be more competitive for a direct
comparison with various experimental findings,
new strategies have to be developed to describe for such
multi-particle interactions within present day transport codes.
A significant first step forward was very recently made;
first results concerning the production of anti-protons
at AGS and SPS energies are quite impressive \cite{Cassing}.
Another strategy could be to exploit microscopically
the concept of two meson doorway states (\ref{twomes})
and their sequential decay
by standard binary scattering processes.



\end{document}